\documentclass[conference]{IEEEtran}
\usepackage{amsmath}
\usepackage[cmintegrals]{newtxmath}
\usepackage{url}
\usepackage{soul,color,graphicx,float,epstopdf}
\usepackage{tablefootnote,textcomp,gensymb}
\usepackage{eurosym,booktabs,multirow}
\usepackage{subcaption}
\usepackage{amsfonts} 
\usepackage{algorithmic}
\usepackage{array}
\usepackage{stfloats}
\usepackage{float}
\usepackage{mathtools}
\usepackage{graphicx}
\usepackage{pdfpages}
\usepackage{balance}
\usepackage{xcolor}
\usepackage{comment}
\usepackage[normalem]{ulem}
\usepackage{dsfont}
\usepackage[linesnumbered, ruled]{algorithm2e}
\usepackage{enumitem}
\usepackage{tikz}
\usetikzlibrary{arrows,shapes,positioning,shadows,trees}
\usepackage[implicit=false]{hyperref}
\usepackage{verbatim}
\usepackage{cite}

\tikzset{
  basic/.style  = {draw, text width=4cm, drop shadow, font=\sffamily, rectangle},
  root/.style   = {basic, rounded corners=2pt, thin, align=center,
                   fill=red!20},
  level 2/.style = {basic, rounded corners=6pt, thin,align=center, fill=blue!20,
                   text width=9em},
  level 3/.style = {basic, thin, align=left, fill=pink!20, text width=8em}
}

\definecolor{Orange}{rgb}{1,0.5,0}

\newcommand*{\rom}[1]{\expandafter\@slowromancap\romannumeral #1@}

\newcommand{\todo}[1]{\textsf{\textbf{\textcolor{Orange}{[#1]}}}}

\begin{document}
\title{Demo: A Digital Twin of the 5G Radio Access Network for Anomaly Detection Functionality}
\author{\IEEEauthorblockN{
Peizheng Li,
Adnan Aijaz,
Tim Farnham,
Sajida Gufran,
Sita Chintalapati
}\\ 
\vspace{-2.00mm}
\IEEEauthorblockA{
\IEEEauthorrefmark{0} Bristol Research and Innovation Laboratory, Toshiba Europe Ltd., U.K.\\
Email: {\{firstname.lastname\}@toshiba-bril.com}
}}

\maketitle

\begin{abstract}
Recently, the concept of digital twins (DTs) has received significant attention within the realm of 5G/6G. This demonstration shows an innovative DT design and implementation framework tailored toward integration within the 5G infrastructure. 
The proposed DT enables near real-time anomaly detection capability pertaining to user connectivity. It empowers the 5G system to proactively execute decisions for resource control and connection restoration. 

\end{abstract}

\begin{IEEEkeywords}
5G, 6G, digital twin, Open RAN, RIC. 
\end{IEEEkeywords}
\section{Introduction}
\subsection{Background and Motivation}
Digital twins (DTs)~\cite{mashaly2021connecting} play a significant role in various domains, including  smart manufacturing, space exploration, energy supply, and smart cities. DT aims to provide a digital copy of a real-world system/process that can be used for low-cost and instant algorithmic development, as well as prototype verification, which can accelerate innovations in all of the involved areas.
DTs have been gaining traction in context of 5G commercialization and 6G research. 
The complexity associated with scaling 5G networks as well as the stringent performance requirements create various challenges for the overall deployment process of 5G. On the other hand, native intelligence capabilities are defining the evolution toward 6G. 
Naturally, DTs become a promising solution. As a virtual replica of the real system, a DT provides the tools for dynamic network optimization, parametric  verifications, network maintenance and diagnostics. While no standardized framework for DTs exist, the wider consensus is that it must provide a high-fidelity representation of the live network, facilitating  seamless integration between cyber and physical domains. 

\begin{figure}[t]   
        \centering   
        \includegraphics[scale=0.46]{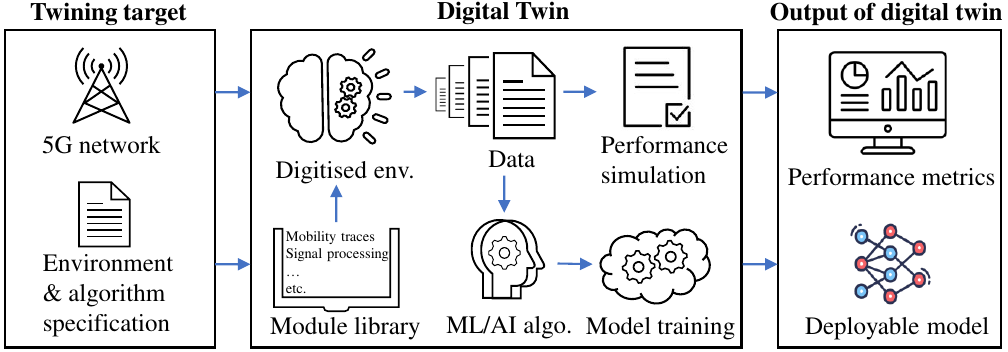}   
            \caption{Typical digital twin functionalities for a 5G system.} \label{fig:dt framework}
            \vspace{-5.00mm}
\end{figure}

There is a plethora of research on digital twins for 5G. Fig.~\ref{fig:dt framework} outlines the key components and functionalities typically considered in a DT for 5G. DT usually has a digitized environment at its core, which approximates real-world entities using computer programs and mathematical models, generating emulated data via third-party libraries~\cite{li2022rlops}. 
While it can communicate with real 5G data for monitoring/diagnostics, it predominantly functions independently as a simulation engine, loosely connected to operational 5G system.
Hence, the effectiveness of a DT  in guiding 5G network operation is constrained due to extended feedback loops and detachment from real-world systems. An external DT incurs higher latency and brings significant information leakage risks.

\subsection{Demonstration Overview}
To address these challenges, we demonstrate a novel approach (illustrated in Fig.~\ref{fig:proposed dt framework}) for DT design and implementation. Our focus is the radio access network (RAN) which handles connectivity for the user equipment (UE). Due to variations in link-level quality and UE density, providing consistent connectivity becomes a challenge. Recent trends toward Open RAN introduce disaggregated components and new interfaces which broaden the threat vector for 5G. This demonstration shows how a DT can be deployed within the 5G RAN to detect anomalies in connectivity. Our DT utilizes key capabilities of Open RAN architecture for integration within the 5G system. The demonstration is based on a near product-grade multi-vendor 5G system, aligned with the O-RAN architecture~\cite{o2021ran}, which includes the radio unit (RU), the distributed unit (DU), the central unit (CU), and the near-real-time RAN intelligent controller (near-RT RIC).


\section{Design and Implementation}
\label{sec: dt}
\subsection{Digital Twin}
\begin{figure*}

  \begin{tabular}[c]{@{}c@{}}
    \begin{subfigure}[c]{.305\linewidth}
      \centering
      \includegraphics[width=\linewidth]{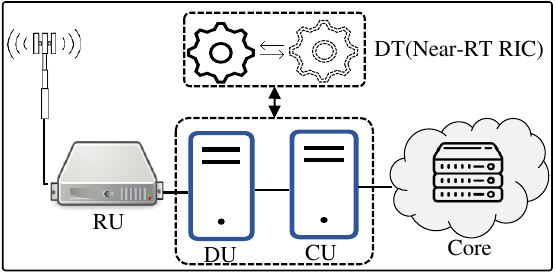}%
      \caption
        {\label{fig:proposed dt framework}%
        }%
        \vspace{-3.25mm}
    \end{subfigure}\\
    \noalign{\bigskip}%
    \begin{subfigure}[c]{.305\linewidth}
      \centering
      \includegraphics[width=\linewidth]{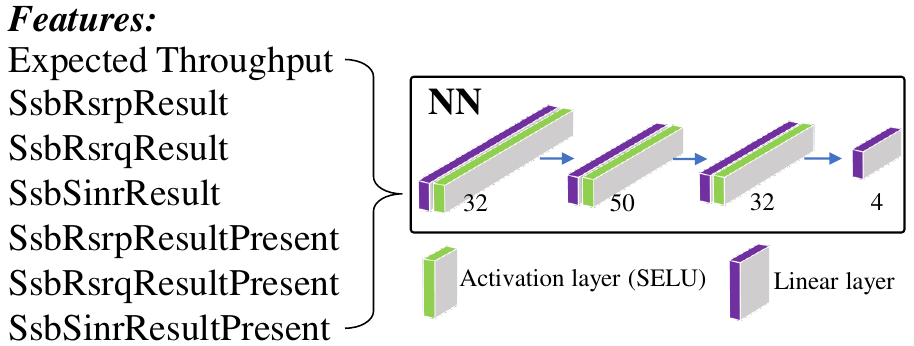}
      \caption
        {
          \label{fig:anomaly detection and NN}%
        }%
    \end{subfigure}
  \end{tabular}
  \hfill
    \begin{subfigure}[c]{.305\linewidth}
    \centering
    \includegraphics[width=\linewidth]{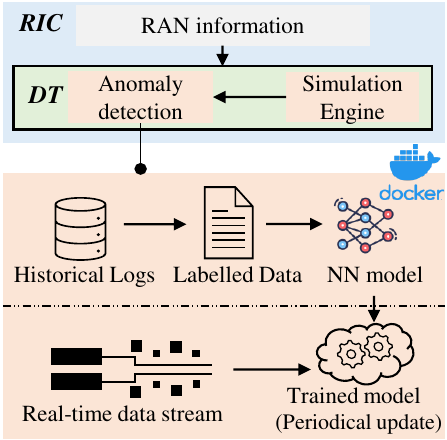}%
    \caption  
      {%
        \label{fig:detailed DT design}%
      }%
  \end{subfigure}
  \hfill
    \begin{subfigure}[c]{.305\linewidth}
    \centering
    \includegraphics[width=\linewidth]{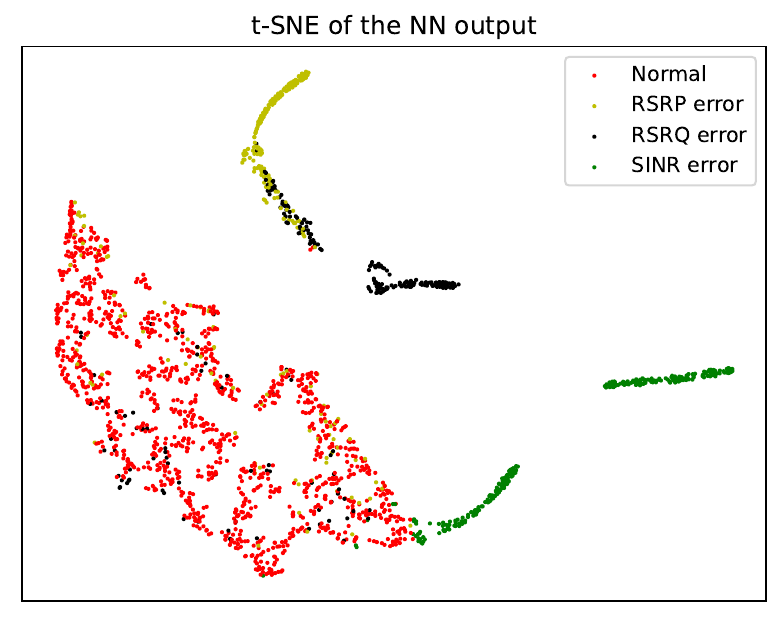}%
    \caption  
      {%
        \label{fig:t-SNE of the NN ouput}%
      }%
  \end{subfigure}

  
  \caption
    {%
    (a) Our proposed digital twin implementation method in 5G RAN; (b) NN architecture for anomaly detection; (c) detailed digital twin functionalities and anomaly detection workflow; (d) t-SNE analysis of the NN outputs on the test set.
      \label{fig:usease1 everything}%
    }
    \vspace{-3.00mm}
\end{figure*}

Our DT design is based on a simulation engine/module, an anamoly detection module, and real-time UE and system information as well as historical logs from different RAN components. 
As shown in Fig.~\ref{fig:detailed DT design}, the simulation engine incorporates channel quality, interference, traffic loading, and the priority of different applications into consideration, to determine the optimal allocation of network resources.  The anomaly detection module is based on  machine learning (ML) techniques. Specifically, neural networks (NNs) are adopted for the detection of compromised connections. The anomaly detection module takes the output of the simulation engine in  final decision-making.
Overall, this DT  resides at the near-RT RIC\footnote{The near-RT-RIC~\cite{o2021ran_ric} is introduced for real-time optimization of the RAN components, enabling more flexible and efficient network operations.} of the 5G Open RAN. 

The DT captures real-time operational information from RAN components via proprietary interfaces implemented by the near-RT RIC vendor. This information includes the number of UEs, and link-level information of each UE in the form of reference signal receive power (RSRP), reference signal received quality (RSRQ), and signal to interference plus noise ratio (SINR). This information is fed to the two modules for optimal scheduling and monitoring UE-level connectivity.


By deploying the DT in the 5G RAN and by exploiting the capabilities of Open RAN, the benefits are threefold: (1) the DT can perform the system performance simulation in real-time (i.e., on the order of 10 msec); (2) it can finish the On-board closed-loop inference and detection of anomalies, which increases the response speed and data security; (3) the DT can interact with other RIC applications for information exchange, fusion and policy control.

\subsection{Anomaly Detection}
The bottom half of Fig.~\ref{fig:detailed DT design} shows anomaly detection workflow in our DT. There are two stages: (1) the NN model generation and training stage; (2) the inference stage that employs the trained NN for prediction with the fuel of the real-time data stream in the near-RT RIC. The model training relies on the prior collected network operation dataset, which is labelled according to the existing abnormal patterns, such as the known errors in RSPR/RSRQ/CQI reports. After deployment, the NN model plays the role of anomaly inference. Note that the trained model is required to be automatically and periodically updated for adapting to the changes in the anomaly patterns of the real 5G system. 
The anomaly detection function is containerized in the docker image, which provides efficient and streamlined deployment and management. The model's entire lifecycle from training and validation to deployment and updating is processed within the container. Meanwhile, Kubernetes is adopted for container orchestration and management among DT and other applications.

Fig.~\ref{fig:anomaly detection and NN} details the  architecture of NN model, depicting input features and output patterns. A fully connected NN is adopted, with an input feature of a $1\times8$ vector encompassing the highlighted parameters,
which are derived from the corresponding UE measurements and system simluations. 
The output of this NN contains $4$ classifications, that is, the normal connectivity, the RSRP error, the RSRQ error and the SINR error. We collected a dataset with a size of $2505$ samples for model training, and the training and testing set is split by $80\%/20\%$ division. Fig.~\ref{fig:t-SNE of the NN ouput} demonstrated the results of applying the t-distributed stochastic neighbour embedding (t-SNE) to the output of the NN on the test set.
In most cases, clear boundaries can be observed for different anomalous patterns. 
The accuracy of anomaly detection is around $90\%$.

The result of the anomaly detection model provides a meaningful indicator of a UE's connection state for the 5G system. Then, the network can take proactive steps to restore or rebuild this UE to network connection. For example, through reallocating physical resources for the affected UE or influencing its association with alternate RUs in the vicinity. These control schemes can be realized by developing specific RIC-centric applications in the 5G Open RAN. 
Eventually, closed-loop 5G network control and optimization can be realized. 

\section{Remarks}
This work demonstrated an integrated DT concept for the 5G RAN which resides at the near-RT RIC level. It is based on a simulation engine, fusion of real and simulated data, and ML techniques. The demo shows DT functionality for anomaly detection for UE-level connectivity management. A short video of the demo is available at:
\url{https://tinyurl.com/Tos08}.

\bibliographystyle{IEEEtran} %

\bibliography{IEEEabrv,references} 

\end{document}